\documentclass[12pt]{article}
\usepackage{epsfig}
\makeatletter
\def\@cite#1#2{({#1\if@tempswa , #2\fi})}
\def\@eqnnum{{\reset@font\rm [\theequation]}}
\makeatother

\begin{document}
\
\vskip 3em

\centerline {\bf Broadband Geodesic Pulses for Three Spin Systems: 
Time-Optimal }
\centerline {\bf  Realization of Effective Trilinear Coupling Terms 
and Indirect SWAP Gates}

\vskip 3em

\centerline {Timo O. Reiss$^1$,
Navin Khaneja$^2$,
Steffen J. Glaser$^1$}
\vskip 3em

$^1$Institut f\"ur Organische Chemie und Biochemie II, Technische 
Universit\"at  M\"unchen,

Lichtenbergstr. 4, 85747 Garching, Germany

$^2$Division of Applied Sciences, Harvard University, Cambridge, MA 02138, USA


\vskip 10em
\begin{abstract}
Broadband implementations of time-optimal geodesic pulse elements are 
introduced
for the efficient creation of effective trilinear coupling terms for 
spin systems
consisting of three weakly coupled spins 1/2. Based on these pulse 
elements, the time-optimal implementation of
indirect SWAP
operations is demonstrated experimentally. The duration of indirect 
SWAP gates based on broadband geodesic
sequence is reduced by  42.3\% compared to
conventional approaches.
\end{abstract}


\vfill \eject

\section{Introduction}

In the absence of relaxation, nuclear magnetic resonance (NMR) 
experiments consist of a sequence of
unitary transformations of the density operator representing the spin 
system of interest \cite{ernst:buch}.  An
important goal of both theoretical and practical interest is the 
design of pulse sequences
that can generate a desired unitary transformation as fast as possible, in
order to reduce losses due to relaxation.  This poses the problem of 
time-optimal control
of quantum systems \cite{navin:toc, navin:qc, Timo:2spin}, which is 
of interest for coherent
spectroscopy in general, as well as for
quantum information processing \cite{sjg:unitary}.

Here, we focus on the realization of effective propagators that 
correspond to the action of an effective Hamiltonian with
trilinear coupling terms which simulate a three-spin interaction in
spin chains consisting of three weakly coupled spins 1/2 (Ising coupling).
In the context of NMR polarization-transfer experiments, methods for 
creating such effective trilininear Hamiltonians have been
developed and used for many years \cite{zzz:ole89, zzz:ole97}.
One approach to create such effective Hamiltonians is based on the
decoupling of certain interactions during the pulse sequence 
\cite{navin:qc, Tseng:trilin, Kim:trilin}.
An earlier, more efficient approach does not rely on decoupling 
\cite{zzz:ole89, zzz:ole97}.
Recently, further improved
sequences \cite{navin:qc} were derived which also avoid decoupling.
Even larger time savings are possible by using {\it geodesic} pulse sequences
\cite{navin:qc} that can be shown to be time-optimal. However, so far 
the geodesic pulse sequences were based on weak rf
pulses which severely limited the range of frequency offsets for 
which the experiment is functional.
  In order to pave the way for practical
applications of these time-optimal pulse elements,  we developed 
broadband versions of the
time-optimal geodesic sequence. Time-optimal indirect SWAP operations
\cite{navin:qc} were realized experimentally in a three-spin system, 
demonstrating the  superior performance of the new sequences.

\section{Theory}

We consider a chain of three heteronuclear spins with coupling 
constants $J_{12}=J_{23}=J$, $J_{13}=0$
and offsets $\nu_1$, $\nu_2$, and $\nu_3$. In a multiple-rotating 
frame \cite{ernst:buch}, the
corresponding free evolution Hamiltonian ${\cal H}_0$ is in general given by

\begin{equation}
{\cal H}_0 = {\cal H}_{c} + {\cal H}_{\it off},
\label{Ham0}
\end{equation}

with the coupling term

\begin{equation}
{\cal H}_{\rm c}= 2 \pi J I_{1z} I_{2z} + 2 \pi J I_{2z} I_{3z}
\label{Ham_coup}
\end{equation}

and the offset term

\begin{equation}
{\cal H}_{\it off}= 2 \pi \nu_1 I_{1z} + 2 \pi \nu_2 I_{2z} + 2 \pi 
\nu_3 I_{3z}.
\label{Ham_offs}
\end{equation}

The same Hamiltonian is valid if e.g. the first and third spins are 
homonuclear and the second spin is
heteronuclear ({\it vide infra}).

Many applications in NMR spectroscopy \cite{zzz:ole89, zzz:ole97} and 
NMR quantum computing \cite{cory:qc, gershenfeld:qc,
bennet:qc} require
unitary transformations of the form
\begin{equation}
{\cal U}_{\alpha \beta \gamma} (\kappa) = \exp \{- {\rm i} \ 2 \pi \ 
\kappa \ I_{1\alpha} I_{2\beta} I_{3\gamma}\},
\label{U_abc}
\end{equation}
  where $\alpha$, $\beta$, $\gamma$ can be $x$, $y$ or $z$.
The time $\tau$ required to realize such a propagator depends on the 
pulse sequence and is a function of $\kappa$ (see Table 1).
The propagator
${\cal U}_{\alpha \beta \gamma} (\kappa)$ can also be expressed as
\begin{equation}
{\cal U}_{\alpha \beta \gamma} (\kappa) = \exp \{- {\rm i} \ \tau 
(\kappa) \ {\cal H}_{\alpha \beta \gamma}\},
\label{U_abcx}
\end{equation}
where ${\cal H}_{\alpha \beta \gamma}$ corresponds to an effective 
trilinear coupling Hamiltonian of the form
\begin{equation}
{\cal H}_{\alpha \beta \gamma} = 2 \pi J_{\it eff}(\kappa) \ 
I_{1\alpha} I_{2\beta} I_{3\gamma},
\label{Ham_abc}
\end{equation}
and the effective trilinear coupling constant $J_{\it eff}(\kappa)$ 
is defined by
\begin{equation}
J_{\it eff}(\kappa)= {{\kappa}\over{\tau (\kappa)}}.
\label{tau}
\end{equation}

For practical applications, $\tau(\kappa)$ should be as short as 
possible and hence the effective coupling constant
$J_{\it eff}(\kappa)$ and the scaling factor
\begin{equation}
s(\kappa)={{J_{\it eff}(\kappa)}\over{J}}={{\kappa}\over{J \ \tau (\kappa)}}
\label{sss}
\end{equation}
should be as large as possible. For $0\leq \kappa\leq 1$,
the theoretical limit $\tau^\ast (\kappa)$ for the minimum time 
required to create a propagator
${\cal U}_{\alpha \beta \gamma} (\kappa)$ is given by \cite{navin:qc}
\begin{equation}
\tau^\ast (\kappa)={\sqrt{\kappa ( 4 - \kappa)}\over{2 J }},
\label{ssc}
\end{equation}
which corresponds to a maximum possible scaling factor
\begin{equation}
s^\ast(\kappa)={{2 \kappa}\over{\sqrt{\kappa ( 4 - \kappa)}}}.
\label{ssw}
\end{equation}
It is sufficient to consider $\tau^\ast (\kappa)$ and $s^\ast 
(\kappa)$ only for $0\leq \kappa\leq 1$, because
$\tau^\ast (2n \pm \kappa) = \tau^\ast (\kappa)$, where $n$ is an 
arbitrary integer \cite{navin:qc}.

\vskip 1em

Schematic pulse sequences corresponding to four different approaches
for the creation of ${\cal U}_{z z z}(\kappa)$
are shown in Fig. 1. These sequences can be further streamlined by 
reducing the number of pulses using well known
rules ({\it vide infra}).
\vskip 1em

{\bf Sequence A} with duration
\begin{equation}
\tau_A(\kappa)=
{{2 + \kappa}\over{2J}}
\end{equation}
is based on the identity \cite{navin:qc}
\begin{equation}
{\cal U}_{zzz} (\kappa)=  {\rm exp}\{  - {\rm i} \pi I_{1z} I_{2x} \}
\ \exp \{- {\rm i} \pi \ \kappa \  I_{2 y} I_{3 z }\}\  {\rm exp}\{ 
{\rm i} \pi I_{1z} I_{2x} \}.
\end{equation}
Equivalent sequences \cite{Tseng:trilin} with the same duration 
$\tau_A(\kappa)$ but with less pulses can be constructed
based on the identity
\begin{equation}
{\cal U}_{zzz} (\kappa)=   V_A
\ \exp \{- {\rm i} \pi \ \kappa \  I_{2 z} I_{3 z }\}\   V_A^{-1}.
\end{equation}
with
\begin{equation}
  V_A= {\rm exp}\{ - {\rm i} {{\pi}\over {2}} I_{2x} \}\
  {\rm exp}\{  - {\rm i} \pi I_{1z} I_{2z} \} \
{\rm exp}\{ - {\rm i} {{\pi}\over {2}} I_{2y} \}.
\end{equation}
Equivalant sequence with the same duration $\tau_A(\kappa)$ can also 
be constructed using CNOT operations
\cite{Kim:trilin}.
\vskip 1em

{\bf Sequence B} with duration
\begin{equation}
\tau_B(\kappa)=
{{1}\over{J}}
\end{equation}
is based on the identity
\cite{zzz:ole89, zzz:ole97}
\begin{equation}
{\cal U}_{zzz} (\kappa)=   V_B
\ \exp \{- {\rm i}\  {{\pi}\over {2}} \ \kappa \  I_{2 x}\}\   V_B^{-1}.
\end{equation}
with
\begin{equation}
  V_B= {\rm exp}\{ - {\rm i}\  {{\pi}\over {2}}\  I_{2y} \} \
  {\rm exp}\{  - {\rm i}\  \pi\  (I_{1z} I_{2z} + I_{2z} I_{3z}) \} .
\end{equation}
\vskip 1em

{\bf  Sequence C} with duration
\begin{equation}
\tau_C(\kappa)=
{{1 + \kappa}\over{2 J}}
\end{equation}
is based on the identity
\cite{navin:qc}
\begin{equation}
{\cal U}_{zzz} (\kappa)=   V_C
\ \exp \{- {\rm i}\  \pi \ \kappa \  (I_{1z} I_{2y} + I_{2y} 
I_{3z})\}\   V_C^{-1}\  {\rm exp}\{  {\rm i}\  {{\pi}\over
{2}}\  \kappa \  I_{2z}
\}.
\end{equation}
with
\begin{equation}
  V_C=
  {\rm exp}\{  - {\rm i}\  {{\pi}\over {2}}  \ (I_{1z} I_{2x} + I_{2x} 
I_{3z}) \} .
\end{equation}
\vskip 1em

Finally, the time-optimal {\bf sequence D} with duration
\begin{equation}
\tau_D(\kappa)=\tau^\ast (\kappa)={\sqrt{\kappa ( 4 - \kappa)}\over{2 J }}
\end{equation}
is based on the identity
\cite{navin:qc}
\begin{equation}
{\cal U}_{zzz} (\kappa)=   V_D \ W  \
\exp \{- {\rm i}  \pi  \sqrt{\kappa ( 4 - \kappa)}  (I_{1z} I_{2z} + 
I_{2z} I_{3z})  +
  {\rm i}  \pi  ( 2 - \kappa)   I_{2x}
\}\   V_D^{-1}
\end{equation}
with
\begin{equation}
  V_D= {\rm exp}\{ - {\rm i}\  {{\pi}\over {2}}\  I_{2y} \}
\end{equation}
and
\begin{equation}
W= {\rm exp}\{ - {\rm i}  \pi (2-{{\kappa}\over{2}}) I_{2x} \}.
\end{equation}
\vskip 1em

The durations $\tau(\kappa)$ and scaling factors
$s(\kappa)=J_{\it eff}(\kappa)/J$ of sequences A-D are shown in Fig. 
2 and are summarized in Table 1. In all four sequences,
$I_2$-selective pulses are required. In addition, $I_1$-selective and 
$I_3$-selective 180$^\circ$ pulses
are required in sequence A  for selective decoupling of $J_{12}$ and 
$J_{23}$ during
parts of the pulse sequence.
In contrast, sequences B-D do not require such $I_1$-selective  or 
$I_3$-selective pulses, which simplifies the experimental
implementation of these pulse sequences if spins $I_1$ and $I_3$ are 
homonuclear. These sequences are also suitable for applications,
where spins $I_1$ and $I_3$ are
equivalent, as in I$_2$S spin  systems \cite{zzz:ole89, zzz:ole97}. 
Furthermore, the fact that decoupling is avoided
in sequences B-D makes these experiments considerably more efficient 
than sequence A ({\it vide infra}). Sequence B is a
straight-forward generalization of a well-known pulse sequence 
developped initially \cite{zzz:ole89, zzz:ole97} for the
special case of
$\kappa=1$. Sequence C, which has been proposed recently
\cite{navin:qc}, is more efficient than sequence B. The geodesic pulse
sequence (sequence D) has  the shortest possible duration for all 
values of $\kappa$
\cite{navin:qc}, see Fig. 2 A (top panel). For $\kappa \rightarrow 0$,
the duration of the geodesic pulse sequence approaches 0, in contrast 
to sequences A-C.
Fig. 2 (middle panel) shows the scaling factors
$s(\kappa)$ and Fig. 2 (bottom panel) shows the relative scaling factors
$s(\kappa)/\kappa$ compared to the scaling factor of sequence B ($s_B=\kappa$).
For $\kappa=1$, the scaling factor $s$ of the geodesic sequence is 
73.2\% larger compared to
sequence A and about 15.5\% larger compared to sequences B and C.
As $\kappa$ approaches 0, the scaling factor $s$ of the geodesic 
pulse sequence becomes infinitely larger than the
scaling factors of  sequences A-C.
For example, for $\kappa=0.01$ the scaling factor $s$ of the geodesic 
sequence is already about a factor of 10 larger
compared to sequences A and B and about a factor of 5 larger compared 
to sequence C.

\smallskip

The basic pulse sequences shown in Fig. 1 only create the desired 
unitary transformations ${\cal U}_{zzz}(\kappa)$ if all
spins are on-resonance in a multiple-rotating frame, i.e. if ${\cal 
H}_{\it off}=0$ (c.f. Eq. 3).
However, for most practical applications, a finite offset range must 
be covered by the pulse sequences.
Broadband versions of sequence B can be found in the literature 
\cite{zzz:ole89, zzz:ole97}.
Broadband versions of sequences A \cite{Tseng:trilin} and C can be 
created in a straight-forward way by inserting additional
$\pi$ pulses in the existing delays to refocus chemical shift evolution.
For example, a broadband version of sequence C is shown in Fig. 3.
The robustness of the broadband sequence with respect to rf 
inhomogeneity and offsets can be further
improved by using the $x$, $-x$,  $-x$, $x$ cycle \cite{MLEV-4} for 
the phases of the four
$\pi$ pulses which are applied to spins
$I_1$ and
$I_3$.

Although no
delays exist in the ideal geodesic pulse sequence shown in Fig. 1D, 
delays can be introduced by replacing the
weak pulse with  amplitude $\nu_{w}= (2-\kappa) J / \sqrt{\kappa ( 4 
- \kappa)}$,
duration $\tau^\ast (\kappa)= \sqrt{\kappa ( 4 - \kappa)}/(2J)$, and
flip angle $\alpha_{w}=\nu_{w} t_{w} 2 \pi =    (2-\kappa) \pi$
by $n$ hard pulses with
flip angle $\alpha_{w}/n$ and $n$
delays of duration
$\Delta=\tau^\ast (\kappa)/n$.
In the limit of $n \rightarrow \infty$, this DANTE-type (Delays 
alternating with Nutations for Tailored Excitation) 
\cite{Morris:Dante}
pulse sequence creates the same effect as the weak pulse and with the 
same limited bandwidth.
For the present application, the DANTE-type sequence approaches the 
ideal sequence if  $\Delta \ll 1/J$.
By inserting $\pi$
pulses in the delays of the DANTE sequence (c.f. Fig. 4), a broadband 
version of the geodesic pulse sequence can be
created. As shown in Fig. 4, the robustness of the broadband geodesic 
sequence with respect to rf inhomogeneity and offset
can be improved by using cycles or supercycles such as $x$, $-x$, 
$-x$, $x$ \cite{MLEV-4} for the phases of each set of
four
$180^\circ$ pulses.

\vskip 1em

In addition to applications in polarization transfer experiments 
\cite{zzz:ole89, zzz:ole97}, propagators corresponding to trilinear
effective coupling terms are useful in the field of quantum 
information processing. For example, so-called $\Lambda_2$ gates
\cite{barenco:gates} can be implemented efficiently based on 
$U_{zzz}(\kappa)$ for $\kappa=1$.
Here, we focus on the implementation of SWAP operations 
\cite{tosch:swap, freeman:swap, bruschweiler:swap} that
make it possible to exchange arbitrary spin states of two spins in a 
coupling network.
For weakly
coupled spins such as in the spin system defined in Eq. (2), a {\it 
direct} SWAP gate
such as ${\it SWAP}(1,2)$ or ${\it SWAP}(2,3)$
  between directly coupled spins $I_1$ and $I_2$, or between $I_2$ and $I_3$
has a minimum duration of \cite{navin:toc}
\begin{equation}
\tau_{{\it SWAP}(1,2)}=\tau_{{\it SWAP}(2,3)}=3/(2 J).
\label{T12}
\end{equation}

An {\it indirect} SWAP
operation ${\it SWAP}(1,3)$ between spins $I_1$ and $I_3$, which are 
not directly coupled, can always
be realized based on the following combination of the direct SWAP 
gates ${\it SWAP}(1,2)$ and ${\it
SWAP} (2,3)$:

\begin{equation}
{{\it SWAP}}(1,3) = {{\it SWAP}}(1,2) \ {{\it SWAP}}(2,3) \   {{\it SWAP}}(1,2)
\label{eq:swap13}
\end{equation}

with an overall duration

\begin{equation}
\tau^{conv}_{{\it SWAP}(1,3)}= 2 \ \tau_{{\it SWAP}(1,2)} + 
\tau_{{\it SWAP}(2,3)} =
9/(2 J).
\label{T13}
\end{equation}

However,  the time-optimal realization of the  indirect SWAP
operation ${\it SWAP}(1,3)$ has a duration of only \cite{navin:qc}

\begin{equation}
\tau^{\ast}_{{\it SWAP}(1,3)}=3 \ \tau^\ast (1)=3 \sqrt{3}/(2 J)
\label{T13opt}
\end{equation}

and hence requires only 57.7 $\%$ of the duration $\tau^{conv}_{{\it 
SWAP}(1,3)}$ of the conventional
sequence (c.f. Eq. 27). This approach is based on the time-optimal 
realization of
propagators ${\cal U}_{\alpha \beta \gamma} (\kappa) $ (c.f.
Eq. 5), which create the desired indirect ${\it SWAP}(1,3)$ gate by 
the following sequence of
operations
\cite{navin:qc}:

\begin{equation}
{\cal U}_{{\it SWAP}(1,3)} = {\cal U}_{zzz}(1) \ {\cal U}_{yzy}(1) \ 
{\cal U}_{xzx}(1) \ \exp({\rm i}
\frac{\pi}{2} I_{2z}).
\label{Uindi}
\end{equation}

Note that all terms in Eq. 29 mutually commute and hence in 
experimental implementations the order of the corresponding pulse 
sequnce elements
is arbitrary.  Based on pulse sequence elements for the realization 
of ${\cal U}_{z z z}(\kappa)$, the sequence of propagators
in Eq. (29) can be realized in a straight-forward way by the pulse 
sequence shown in Fig.\
5 C. The final $90^\circ_{-z}$ rotations can either be implemented by 
a composite pulse
such as
$90^\circ_x \  90^\circ_y \  90^\circ_{-x} $or by adjusting the 
phases of all following
pulses and of the receiver
\cite{5qubits}. In general,
$z$ rotations (by angle $\varphi$) can be implemented by an 
additional phase shift (by angle $- \varphi$) of all
following r.f.\ pulses that are applied to this spin and of the
receiver phase for this spin \cite{5qubits}.

\section{Experiments}

In order to test the performance of the new geodesic pulse sequences, 
we used the spin
system of the amino moiety of [$^{15}$N]-acetamide as a  model system 
(see Fig. 6) that
corresponds closely to the model Hamiltonian
${\cal H}_0= {\cal H}_{c} + {\cal H}_{\it off}$ defined in Eqs. (1-3).
[$^{15}$N]-acetamide (Chemotrade GmbH) was dissolved in DMSO-d6 and 
all measurements were
performed on a Bruker 600 MHz DMX
spectrometer (Bruker Analytik GmbH) at a temperature of 298 K.
Here, spins $I_1$ and $I_3$ correspond to the amino protons, whereas 
$I_2$ corresponds to the $^{15}$N spin with
$J_{12}=$ 88.8 Hz $\approx J_{23}=$ 87.3 Hz $\gg J_{13}=2.9$ Hz. 
Additional $^4 J(^1H, ^{1}H)$ and $^3 J(^1H,
^{15}N)$couplings (0.7 Hz and 1.2 Hz) of $I_1$ and $I_2$ to the 
methyl protons of [$^{15}$N]-acetamide  are about two
orders of magnitude smaller than the
$^1 J(^1H, ^{15}N)$ couplings.

Compared to a fully heteronuclear spin system, the
relatively small frequency difference $\Delta \nu_{13}= 358 $ Hz of 
the amino protons (spins $I_1$ and $I_3$) makes it difficult
to apply short selective pulses to spin $I_1$ that do not affect spin 
$I_3$ (and vice versa) as required in sequence A (and also
for its broadband implementation using additional refocussing 
pulses). In our experiments, we implemented spin-selective proton
pulses by a combination of hard pulses and delays. For example, if 
spin $I_1$ is irradiated on resonance, a selective
180$^\circ_x(I_1)$ pulse can be implemented by the pulse sequence element

\hskip 3cm 90$^\circ_x(I_1, I_3)$  \ \ -\ \ $\delta$\ \ - \ \ 
180$^\circ_x(I_2)$    \ \ -\ \ $\delta$\ \ - \ \
180$^\circ_x(I_2)$  90$^\circ_x(I_1, I_3)$,

where $\delta=2/(4 \Delta \nu_{13})=698$ $\mu$s. Similarly, a 
selective 180$^\circ_x(I_3)$ pulse can be implemented by the same 
pulse
sequence element if the last 90$^\circ_x(I_1, I_3)$ pulse is replaced 
by 90$^\circ_{-x}(I_1, I_3)$.
\smallskip

Based on broadband versions of sequences A, C, and D, we implemented 
the sequence shown in
Fig. 5 C, which  realizes a ${\it SWAP}(1,3)$ operation for 
$\kappa=1$. Note that the sequence
of Fig. 5 C still allows for a variation of
$\kappa$, which makes it possible to test the theoretically expected 
$\kappa$ dependence of the sequences ({\it vide
infra}).
For $\kappa = 1$, the sequences for the ${\it SWAP}(1,3)$ gate were 
successfully tested for a large
number of initial states of the spin system.   Three illustrative 
examples are presented in Fig. 7,
where
  $^1$H spectra of the amino protons (spins $I_1$ and $I_3$) of 
[$^{15}$N]-acetamide are shown. Left (A-C) and right 
(A$^\prime$-C$^\prime$)
spectra reflect the states before and after an indirect ${\it 
SWAP}(1,3)$ operation, repectively. The
initial spin states were prepared to be
$\rho(0)=I_{1x}$ (c.f. Fig. 7A),  (B) $\rho(0)=2I_{1x} I_{2z}$ (c.f. 
Fig.\ 7 B), and (C) $\rho(0)=I_{1x} + 2 I_{2z} I_{3x}$ (c.f.
Fig.\ 7 C). As expected, the states of the two proton spins (spins 
$I_1$ and $I_3$) are swapped for
arbitrary initial states.

In order to compare the durations and $\kappa$ dependence of the 
indirect ${\it SWAP}(1,3)$ sequences,
we  measured  the efficiency of inphase transfer from $I_{1x}$ to 
$I_{3x}$ (c.f. Fig. 7 A): For an
initial density operator of
$\rho(0)=I_{1x}$, we defined the
transfer efficiency $\eta_{13}(\tau)$ for a given pulse sequence of 
duration $\tau$ as

\begin{equation}
\eta_{13}(\tau)= {{\langle I_{3x} \rangle (\tau)} \over{ \langle 
I_{1x} \rangle (0)}},
\label{defeta}
\end{equation}

where $\langle I_{1x} \rangle (0)$ is the initial expectation value 
of $I_{1x}$ and
$\langle I_{3x} \rangle (\tau)$ is the expectation value of $I_{3x}$ 
after the pulse sequence.
The corresponding experimental values of $\eta$ were determined by 
dividing the integral of the
  spin $I_3$ multiplet in the final spectrum by the integral of the
  spin $I_1$ multiplet in the initial spectrum.

Fig. 8 summarizes the theoretical and experimental curves of the 
transfer efficiency $\eta_{13}(\tau)$ based on broadband versions of
sequences, A, C, and D.
In the experiments, the parameter $\kappa$ (c.f. Eq. 4) was varied in 
the range $0 \leq \kappa\leq 2$. Fig. 8 A shows the theoretical
$\tau$ dependence of the transfer efficiency
$\eta_{13}$ for
  the pulse sequences of Fig.\ 1 A, C, and D, assuming ideal 
spin-selective hard pulses
without rf inhohogeneity and an isolated, ideal three-spin system. 
All spins are assumed
to be on-resonance (${\cal H}_{\it off}=0$) in a multiple rotating frame with
$J_{12}=J_{23}= 88$ Hz and $J_{13}= 0$ Hz.
As expected (c.f. Table 1), transfer efficiencies of $\eta_{13}=1$ 
(corresponding to a complete SWAP operation) are found for $\tau= 
51.1$
ms (sequence A),
$34.1$ ms (sequence C), and
$29.5$ ms (sequence D).

More realistic values of the transfer efficiency $\eta_{13}(\tau)$ to 
be expected for our model system
were obtained by simulating the time
evolution of the density operator during the broadband pulse 
sequences for the actual coupling network (using the experimentally
determined coupling constants and frequency offsets) and taking into 
account experimental pulse sequence parameters (see Fig.\
8 B). Nominal rf amplitudes of 35.7 kHz and 5.5 kHz were assumed for 
$^1$H and $^{15}$N pulses,
respectively (corresponding to 90$^\circ$ pulse durations of 7 $\mu$s 
and 45 $\mu$s).
The effects of rf inhomogeneity were taken into account by assuming a 
Gaussian distribution of the rf amplitudes with a full width at half
hight of 10$\%$
\cite{sjg:tocsy}. Relaxation effects were not included.
The
simulated  $\eta_{13}(\tau)$ curves in Fig. 8 B qualitatively match 
the ideal curves shown in Fig.\ 8 A. In particular, the position of
the maxima appear at very similar pulse sequence durations $\tau$. 
However, the amplitude of the $\eta_{13}(\tau)$ curves is
decreased due to the effects of experimental imperfections.

In Fig.\  8 C, experimentally determined transfer efficiencies 
$\eta_{13}(\tau)$ are shown for the three pulse sequences. A
reasonable match is found between experimental and simulated curves. 
The experimentally determined bandwidth
covered by the broadband geodesic sequence was about 3.5 kHz for $^1$H and  2.5
kHz for
$^{15}$N for the given pulse sequence parameters.

\section{Discussion}

Broadband versions of a new class of pulse sequences for the 
simulation of trilinear coupling terms were developed.
Using the amino group of [$^{15}$N]-acetamide as a
model system, the theoretically predicted properties \cite{navin:qc} 
of the new sequence C and of the time-optimal geodesic
sequence D were verified and efficient exchange of the spin state of 
indirectly coupled spins was demonstrated.
It is expected that the new broadband pulse sequences will find 
applications both in  quantum
information processing and in coherent spectroscopy.

\vskip 3em
{\bf Acknowledgments}

This work was supported by the DFG under grants Gl 203/4-1 and 4-2. 
N. K. would like to thank Darpa grant 496020-0101-00556
and NSF Qubic grant 0218411.

\newpage

\begin{table*}
\caption{Pulse sequence durations $\tau$  and scaling factors 
$s=J_{\it eff}(\kappa)/J$ of
effective  trilinear coupling constants.}
\begin{center}
\begin{tabular}{ccccc} \hline
\  & A \ \ \ \ & \ \ \ \ B \ \ \ \  &C \ \ \ \  &D \ \ \ \  \\ \hline
$\tau(\kappa)$ &   ${{2 + \kappa}\over{2 J}}$ &${{1}\over{J}}$ & 
${{1 + \kappa}\over{2 J}}$   &
${{\sqrt{\kappa(4-\kappa )}}\over{ 2 J}}$
\\
$\tau(1)$ &   ${{1.5}\over{J}}$ &   ${{1}\over{J}}$& 
${{1}\over{J}}$  &   ${{\sqrt{3}}\over{2 J}}\approx
{{0.866}\over{J}}$  \\
$s(\kappa)$ &  ${{2 \kappa }\over{2 + \kappa}}$  &   $\kappa$  & 
${{2 \kappa }\over{
1 + \kappa}}$   &    ${{2 \kappa }\over{\sqrt{\kappa(4-\kappa )}}}$ \\
$s(1)$ &  ${{2}\over{3}}\approx 0.666$  &  $1$  &  $1$   &   ${{2 
}\over{\sqrt{3}}}\approx1.155$  \\
$\tau_{{\it SWAP}(1,3)}(J)$ &   ${{4.5}\over{J}}$ & 
${{3}\over{J}}$&   ${{3}\over{J}}$  &   ${{3
\sqrt{3}}\over{2 J}}\approx{{2.598}\over{J}} $\\
$\tau_{{\it SWAP}(1,3)}(88 {\rm Hz})$ &  51.1 ms  &  34.1 ms &  34.1 
ms   &   29.5 ms  \\
\hline
\end{tabular}
\label{tab:times}
\end{center}
\end{table*}
\

\begin{figure}[p]
\begin{center}
\epsfig{file=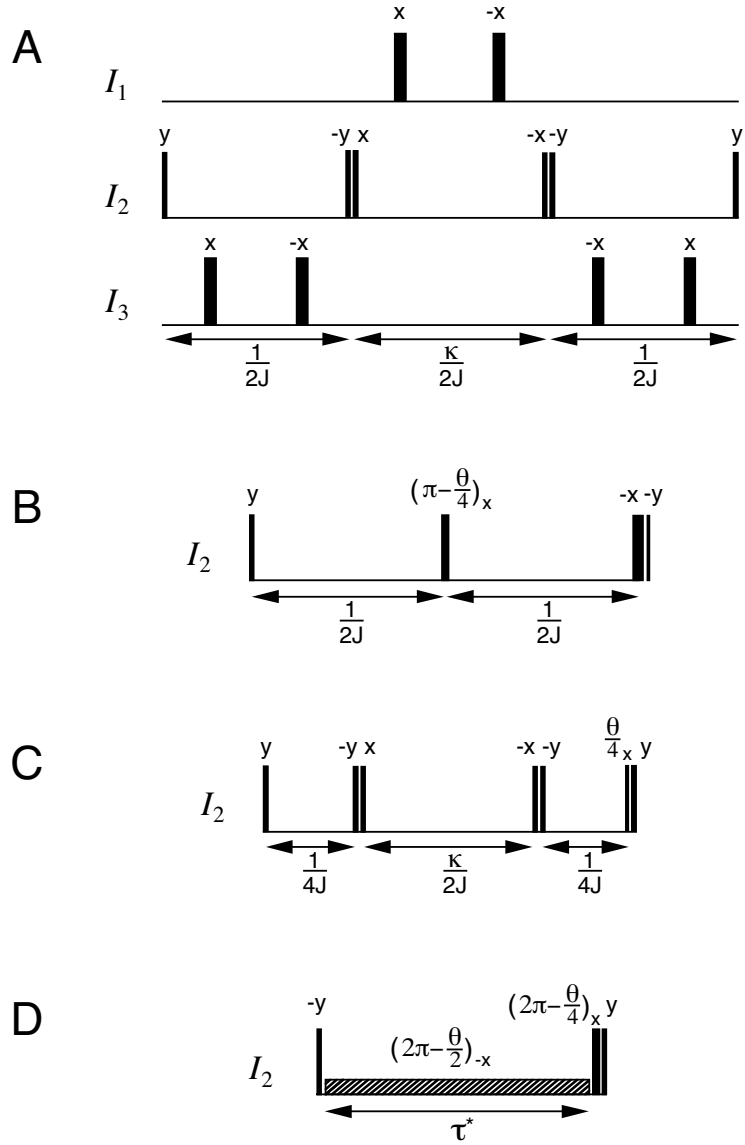, scale=0.8, angle=0, clip=}
\end{center}
\caption[]{Schematic representation of four basic (narrowband) pulse sequences 
for the creation of a propagator
${\cal U}_{zzz} (\kappa) = \exp \{- {\rm i} \ \theta \ I_{1z} I_{2z} 
I_{3z}\}$ with $\theta=2 \pi \kappa$.
If not explicitely specified otherwise,
narrow and wide vertical bars represent spin-selective $\pi/2$ and 
$\pi$ pulses, respectively.
(A) Example of a pulse sequence based on selective decoupling 
\cite{navin:qc, Tseng:trilin, Kim:trilin}, (B) conventional
pulse sequence without decoupling \cite{zzz:ole89, zzz:ole97}, (C) improved
sequence without decoupling \cite{navin:qc}, (D) time-optimal 
geodesic pulse sequence with $\tau^\ast=\sqrt{\kappa ( 4 -
\kappa)}/2 J$  \cite{navin:qc}.}
\label{Fig1}
\end{figure}

\begin{figure}[p]
\begin{center}
\epsfig{file=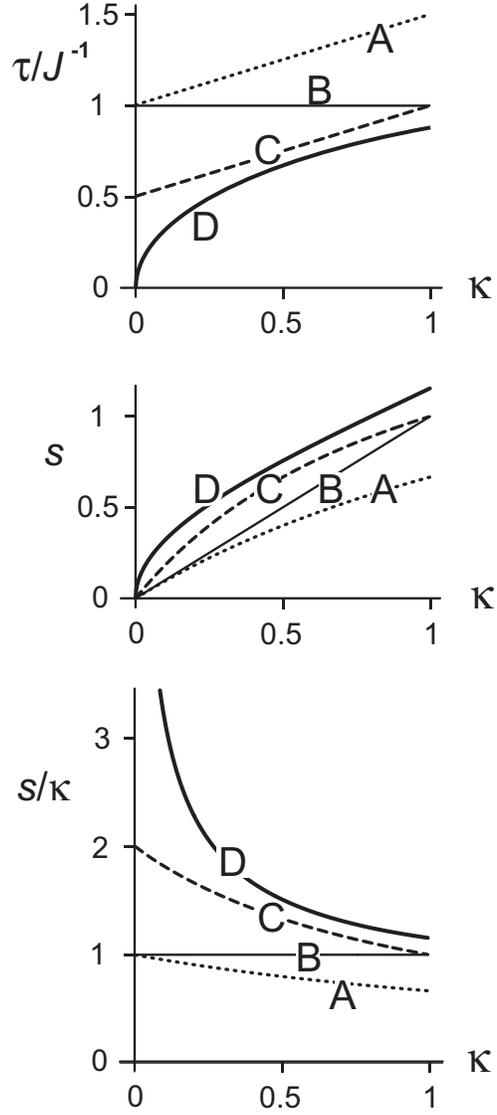, scale=1, angle=0, clip=}
\end{center}
\caption[]{Durations $\tau (\kappa)$ (top panel), scaling factors 
$s=J_{\it eff}(\kappa)/J=\kappa/J
\tau (\kappa)$ (middle panel), and relative scaling factors 
$s/s_B=s/\kappa$ (bottom panel) of the four
basic pulse sequences A-D shown in Fig. \ref{Fig1}.}
\label{Fig2}
\end{figure}

\begin{figure}[p]
\begin{center}
\epsfig{file=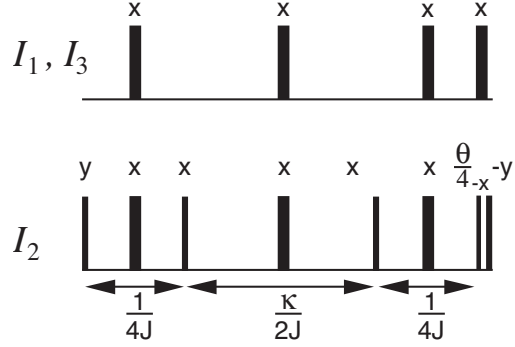, scale=0.8, angle=0, clip=}
\end{center}
\caption[]{Broadband version of sequence C shown in Fig. \ref{Fig1}.}
\label{Fig3}
\end{figure}

\begin{figure}[p]
\begin{center}
\epsfig{file=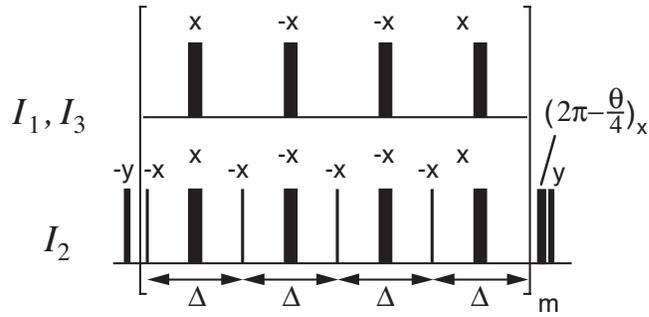, scale=0.8, angle=0, clip=}
\end{center}
\caption[]{Broadband version of the geodesic sequence shown in Fig. 1D
for the time-optimal implementation of ${\cal U}_{zzz} (\kappa)$.
The pulse sequence element in brackets has a duration of 4$\Delta$ 
and is repeated $m$ times.
The narrow vertical bars in the bracket correspond to hard pulses 
with flip angles
$2 \pi \nu_w \tau^\ast/n$ and $\Delta=\tau^\ast/n$ with $n=4m$.}
\label{Fig4}
\end{figure}

\begin{figure}[p]
\begin{center}
\epsfig{file=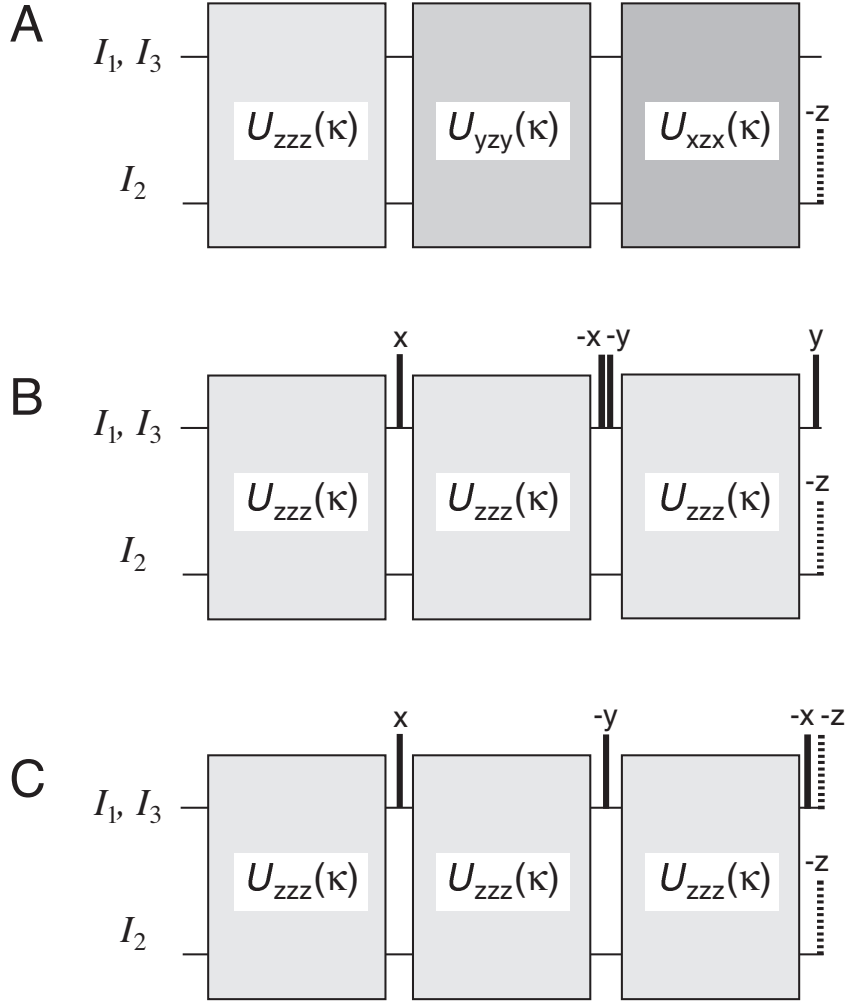, scale=0.8, angle=0, clip=}
\end{center}
\caption[]{Pulse sequences implementing the  indirect SWAP operation ${\cal 
U}_{{\it SWAP}(1,3)}$ for $\kappa=1$.
(A) Schematic implementation according to  Eq. 29, (B) equivalent 
implementation based on pulse
sequence elements (c.f. Fig. 1) that create the propagator ${\cal 
U}_{zzz} (\kappa)$, (C) streamlined
pulse sequence with a minimum number of $90^\circ$ pulses.}
\label{Fig5}
\end{figure}

\begin{figure}[p]
\begin{center}
\epsfig{file=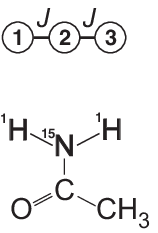, scale=2, angle=0, clip=}
\end{center}
\caption[]{The model system (top) consisting of a chain of three 
coupled spins 1/2 with $J_{12}=J_{23}=J$ and
$J_{13}=0$
  is approximated by the spins of the amino moiety (printed in 
boldface) of [$^{15}$N]-acetamide (bottom).}
\label{Fig6}
\end{figure}

\begin{figure}[p]
\begin{center}
\epsfig{file=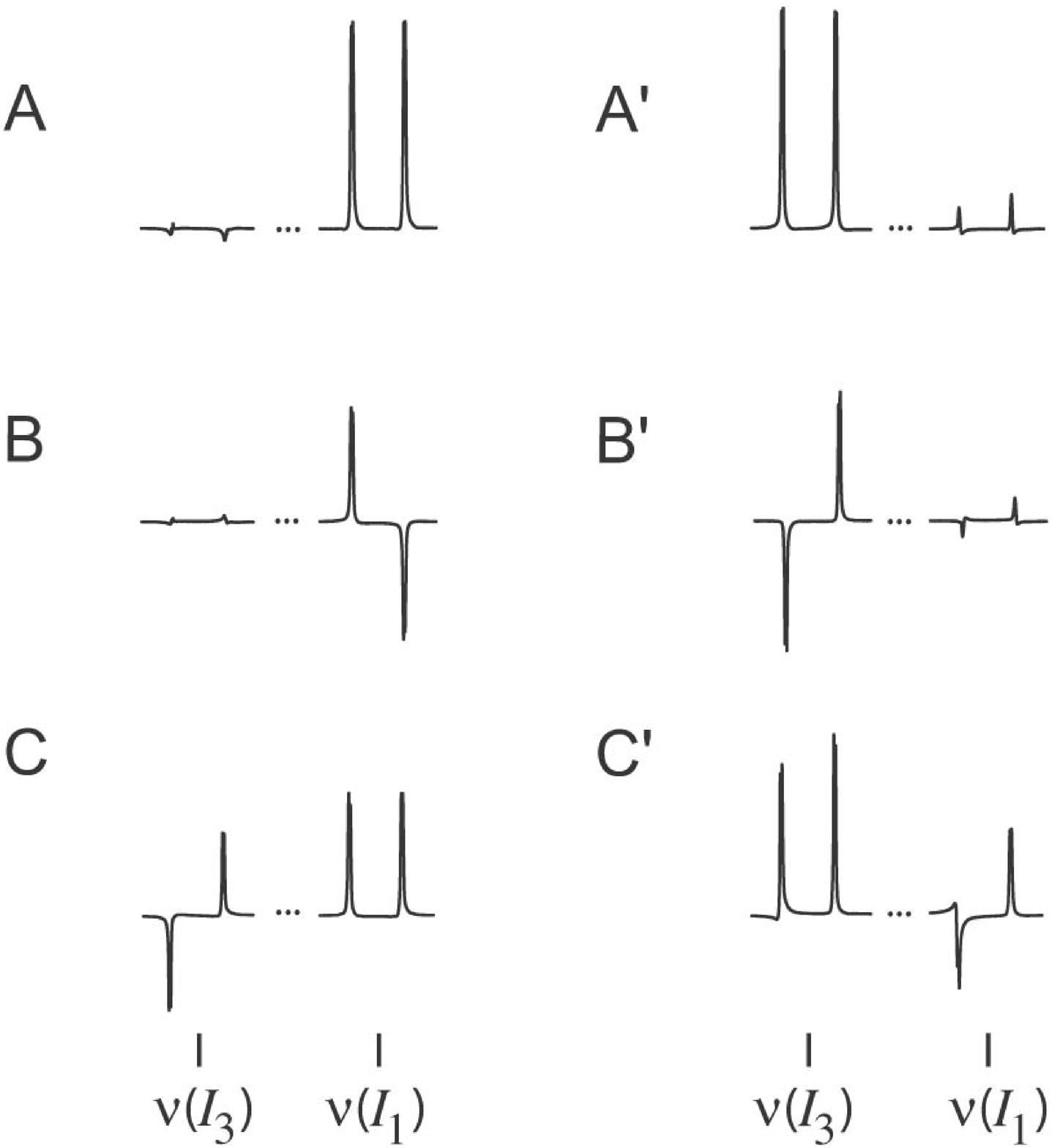, scale=0.4, angle=0, clip=}
\end{center}
\caption[]{$^1$H spectra of the amino protons (spins $I_1$ and $I_3$) of 
[$^{15}$N]-acetamide before (A-C) and after
(A$^\prime$-C$^\prime$) an indirect ${\it SWAP}(1,3)$ operation based 
on Fig. 5 C and the broadband
sequence shown in Fig. 3 for the creation of
${\cal U}_{zzz} (\kappa)$ for $\kappa=1$. The initial spin states 
were prepared to be (A)
$\rho(0)=I_{1x}$,  (B) $\rho(0)=2I_{1x} I_{2z}$, and (C) 
$\rho(0)=I_{1x} + 2 I_{2z} I_{3x}$, respectively.}
\label{Fig7}
\end{figure}

\begin{figure}[p]
\begin{center}
\epsfig{file=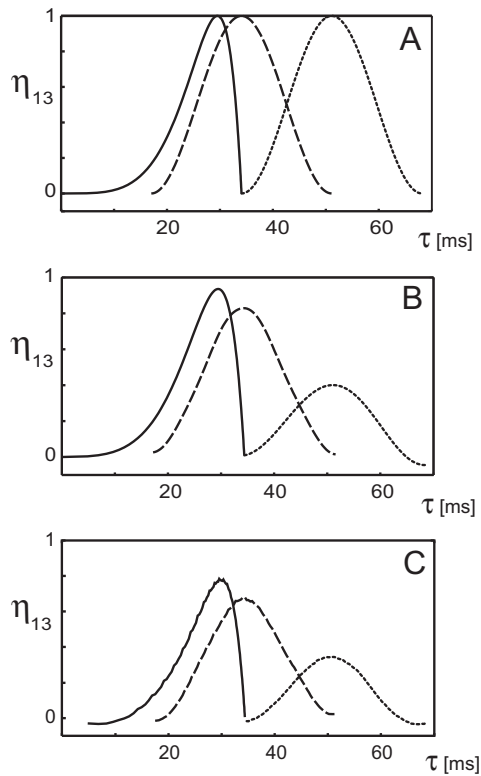, scale=0.6, angle=0, clip=}
\end{center}
\caption[]{Transfer efficiency $\eta_{13}(\tau)$ (c.f. Eq. 30)
based on broadband versions of sequences, A (dotted curves), C 
(dashed curves), and D (solid curves).
(A) Theoretical curves assuming an ideal spin system (see Fig. 6 top) 
and ideal rf pulses, (B) simulations based on the
coupling constants of [$^{15}$N]-acetamide and assuming finite pulse 
durations and realistic rf
inhomogeneity (see text), (C) experimental transfer curves.}
\label{Fig8}
\end{figure}

\end{document}